\documentclass{ws-ijmpa}
\usepackage[super,compress]{cite}
\usepackage{xcolor}

\newcommand\numberthis{\addtocounter{equation}{1}\tag{\theequation}}

\newcommand*{\Scale}[2][4]{\scalebox{#1}{\ensuremath{#2}}}%

\newcommand{\black}{\color{black}}

\begin{document}
\markboth{T. Helpin $\&$ M. S. Volkov}{A metric-affine version of the Horndeski theory}

%
\catchline{}{}{}{}{}
%

\title{A metric-affine version of the Horndeski theory
}

\author{Thomas Helpin}

\address{Institut Denis Poisson, UMR - CNRS 7013,\\
Universit\'{e} de Tours, Parc de Grandmont, 37200 Tours, France\\
thomas.helpin@lmpt.univ-tours.fr}

\author{Mikhail~S.~Volkov}

\address{Institut Denis Poisson, UMR - CNRS 7013, \\ 
	Universit\'{e} de Tours, Parc de Grandmont, 37200 Tours, France\\
	and\\
	Department of General Relativity and Gravitation, Institute of Physics,\\
	Kazan Federal University, Kremlevskaya street 18, 420008 Kazan, Russia\\
michael.volkov@idpoisson.fr}

\maketitle

\begin{history}
\received{Day Month Year}
\revised{Day Month Year}
\end{history}

\begin{abstract}
	
	\noindent
	 We study  the metric-affine  versions of scalar-tensor theories whose   connection enters the action only algebraically. 	 
	 We show that 
the connection  can be integrated out in this case, resulting in an equivalent  metric theory. Specifically, 
	 we consider the metric-affine generalisations of the  subset of the Horndeski theory  whose action  is linear in second derivatives of the scalar field. 
	 We determine the connection and find that it can describe 
	a \textit{scalar-tensor} Weyl geometry without  a Riemannian frame. 
	Still, as this connection enters the action  algebraically,  the theory  
	admits the dynamically  equivalent (pseudo)-Riemannian formulation in the form of 
	an effective metric theory with  an extra K-essence  term. This  may have  interesting phenomenological applications.
\end{abstract}

\ccode{PACS numbers: 04.20.-q, 04.50.Kd}

\section{Introduction}

Within the framework of scalar-tensor theories of Horndeski\cite{Horndeski1974} and  beyond Horndeski \cite{PhysRevD.89.064046,Gleyzes:2014dya}  and in the Degenerate Higher Order Scalar-Tensor (DHOST) theories\cite{Langlois_2016,BenAchour:2016fzp} one assumes the connection to be Levi-Civita. 
All of these theories  evade  the  Ostrogradsky ghost.\cite{Woodard:2015zca} 
The Horndeski theory is the most general scalar-tensor  theory with second order equations of motion. Recently it was rediscovered via the covariantization of the Galileon \cite{Nicolis:2008in,Deffayet:2009wt} and was found to be equivalent to the Generalized Galileon theories \cite{Deffayet:2011gz}. 

In this text we consider  the  MAG (Metric-Affine Gravity) generalisations of the Horndeski theory. In these  theories the gravitational interaction is encoded in  two \textit{a priori} independent fields: the space-time metric $g_{\mu\nu}$ and the distortion tensor $C^\alpha_{\ \mu\nu}$. The latter characterizes the deviation of the 
independent  connection $\Gamma^{\alpha}_{\ \mu\nu}=\bigl \lbrace ^\alpha_{\ \mu\nu}\bigr\rbrace_g+C^\alpha_{\ \mu\nu}$ from the Levi-Civita connection $\bigl \lbrace ^\alpha_{\ \mu\nu}\bigr\rbrace_g$ associated with the spacetime metric. The non-Riemannian part of the connection\footnote{The Riemannian structure of a smooth manifold is related to the metric only, and hence the geometry of the metric-affine theory is always (pseudo-)Riemannian. The  ``non-Riemann" nature of the theory lies in the assumption that in addition to the Levi-Civita connection there may exist a different gravitational connection which couples to the matter and which should be determined via the least action principle.} may have two independent origins: the torsion $T^{\alpha}_{\ \mu\nu}= 2\Gamma^{\alpha}_{\ [\mu\nu]}$ and the  non-metricity $Q_{\alpha\mu\nu}=\overset{\scriptscriptstyle \Gamma}\nabla_\alpha g_{\mu\nu}$. In the MAG formalism with both torsion and non-metricity the Einstein-Hilbert action enjoys the projective gauge symmetry associated with the transformations $\Gamma^{\alpha}_{\ \mu\nu}\to\Gamma^{\alpha}_{\ \mu\nu}+\xi_\mu\delta{^\alpha_\nu}$. This symmetry acts on the vectorial part of the connection, and for consistency reasons\cite{Hehl1978} any metric-affine theory containing the Einstein-Hilbert term must be\footnote{It is still possible to explicitly break the projective symmetry of metric-affine scalar tensor actions with Lagrange multipliers. \cite{Li_2012}}  projectively invariant. There is currently a growing interest in the metric-affine scalar-tensor theories \cite{Li_2012,PhysRevD.100.064018,GALTSOV2019453,Helpin:2019kcq,PhysRevD.96.084023,Aoki:2018lwx,PhysRevD.99.104020,PhysRevD.100.044037,Shimada:2018lnm}. {\black In most of the previously studied cases}
the Levi-Civita covariant  derivatives of the scalar field have been replaced by the  covariant derivatives with respect to the independent connection in the ``minimal" way:
$\overset{\scriptscriptstyle g}\nabla{_\mu}\overset{\scriptscriptstyle g}\nabla_\nu\phi\to\overset{\scriptscriptstyle \Gamma}\nabla{_\mu}\overset{\scriptscriptstyle \Gamma}\nabla_\nu\phi$.
We wish to study the effect of relaxing this minimal  prescription. 
\section{General Setting}
The Ricci tensor associated to the independent connection can be decomposed as $\overset{\scriptscriptstyle \Gamma}R_{\mu\nu}=\overset{\scriptscriptstyle g}{R}_{\mu\nu}+\mathcal{N}_{\mu\nu}$ where 
\begin{equation}
\mathcal{N}_{\mu\nu}=\overset{\scriptscriptstyle g}\nabla_\alpha C^\alpha_{\ \nu\mu}-\overset{\scriptscriptstyle g}\nabla_{\nu}C^\alpha_{\ \mu\alpha}+C^\alpha_{\ \alpha\lambda}C^\lambda_{\ \nu\mu}-C^\alpha_{\ \nu\lambda}C^\lambda_{\ \mu\alpha}
\end{equation}
encodes the contribution of the non-Riemannian part of the connection. Similarly, the {\black Lagrangian} of any standard metric-affine scalar-tensor theory 
$\mathcal{L}(g,\Gamma,\partial\Gamma,\phi,\nabla\phi,\ldots, (\overset{\scriptscriptstyle \Gamma}\nabla\nabla\phi)^n)$,
where $n$ is the maximal power of second derivatives, 
 can be decomposed into the metric part $\mathcal{L}_g$  and the distortion part $\mathcal{L}_{C}$: 
\begin{eqnarray}\label{MAGproj}
\mathcal{L}&=&\mathcal{L}_g(g,\partial g,\partial\partial g,\phi,\nabla\phi,\ldots ,(\overset{\scriptscriptstyle g}\nabla\nabla \phi)^n) \\ \nonumber 
&&+\mathcal{L}_C(g,\phi,\nabla\phi,\ldots, (\overset{\scriptscriptstyle g}\nabla\nabla \phi)^{n-1},C,\ldots, C^n,\overset{\scriptscriptstyle g}\nabla C).
\end{eqnarray}
When the action contains  only  terms linear  in curvature and when they  do not couple to 
 the second derivatives  $\overset{\scriptscriptstyle \Gamma}\nabla\nabla \phi $, then 
 the distortion enters the action solely algebraically. In this case it can be treated as an auxiliary field. The on-shell value of the distortion tensor  will depend on $(g,\phi,\nabla\phi,\ldots,(\overset{\scriptscriptstyle g}\nabla\nabla\phi)^{n-1})$ and can be integrated out from the action resulting in a dynamically equivalent  metric action. Let us clarify this  statement by showing that the equations of motion for the scalar field are indeed equivalent  in the metric-affine and effective metric descriptions, at least if $n=1$. 
 Let $S_{eff}[g,\phi]=S_g[g,\phi]+\tilde{S}_C[g,\phi,C(g,\phi,\nabla \phi)]$ denote the effective metric theory obtained after integrating out the connection. Varying  $S$ and $S_{eff}$ with respect to the scalar field yields 
\begin{eqnarray}
E_\phi&=&\dfrac{1}{\sqrt{-g}}\dfrac{\delta S}{\delta\phi}=\dfrac{1}{\sqrt{-g}}\left(\dfrac{\delta S_g}{\delta \phi}+\dfrac{\delta S_C}{\delta \phi}\right),\qquad  \nonumber \\
\tilde{E}_\phi&=&\dfrac{1}{\sqrt{-g}}\dfrac{\delta S_{eff}}{\delta\phi}=\dfrac{1}{\sqrt{-g}}\left(\dfrac{\delta S_g}{\delta \phi}+\dfrac{\delta \tilde{S}_C}{\delta \phi}\right).
\end{eqnarray}
By definition, 
\begin{equation}\label{EqPhiMA}
\dfrac{\delta S_C}{\delta \phi}=\Scale[0.9]{\sqrt{-g}}\dfrac{\partial \mathcal{L}_C}{\partial \phi}-\partial_\alpha \dfrac{\partial (\Scale[0.9]{\sqrt{-g}}\mathcal{L}_C)}{\partial \partial_\alpha \phi}.
\end{equation}
Let us emphasize that  the distortion in $\mathcal{L}_C(g,\phi,\nabla\phi,C)$ is considered as an independent variable, whereas in $\tilde{\mathcal{L}}_C(g,\phi,\nabla\phi,C(g,\phi,\nabla \phi))$ it 
is a function of $\phi$ and $\nabla\phi$. Using the chain rule one has, 
\begin{equation}
\dfrac{\delta \tilde{S}_C}{\delta \phi}=\Scale[0.87]{\sqrt{-g}}\dfrac{\partial \tilde{\mathcal{L}}_C}{\partial \phi}\bigg|_C-\partial_\alpha \dfrac{\partial (\Scale[0.87]{\sqrt{-g}}\tilde{\mathcal{L}}_C)}{\partial \partial_\alpha \phi}\bigg|_{C}+\Scale[0.87]{\sqrt{-g}}\dfrac{\partial \tilde{\mathcal{L}}_C}{\partial C^\alpha_{\ \mu\nu}}\dfrac{\partial C^\alpha_{\ \mu\nu}}{\partial \phi}-\partial_\alpha\Big(\dfrac{\partial( \Scale[0.87]{\sqrt{-g}}\tilde{\mathcal{L}}_C)}{\partial C^\beta_{\ \mu\nu}}\dfrac{\partial C^\beta_{\ \mu\nu}}{\partial \partial_\alpha\phi}\Big).
\end{equation}
When the distortion tensor appears in the action algebraically, 
its field equation  is equivalent to $\Scale[0.9]{\dfrac{\partial\tilde{\mathcal{L}}_C}{\partial C^\alpha_{\ \mu\nu}}=0}$. Therefore we 
conclude that the equations of motion for the scalar field are the same in both cases:  $E_\phi=\tilde{E}_\phi$. 
This proof generalizes to the equations for the metric, and also to any theory with $n>1$ whose action  is algebraic in the connection. 
On the other hand, theories where the second derivatives of the scalar field couple to the curvature terms give rise {\black to first order} differential equations for the distortion tensor. For example, 
any theory  whose Lagrangian contains  the term 
\begin{equation}
G_5(\phi,X)\overset{\scriptscriptstyle \Gamma}G{}^{\alpha\beta}\overset{\scriptscriptstyle \Gamma}\nabla_\alpha\nabla_\beta\phi
\end{equation}
should support a dynamical connection. Whether or not such theories can be ghost-free is an interesting question, which is however 
beyond the scope of this paper. We shall now fix 
the definitions and notation needed in order to solve and analyze the field equations for the distortion tensor. 

The Palatini tensor is the variation of the Einstein-Hilbert term with respect to the connection, 
\begin{equation}
P_\alpha^{\ \mu\nu}=\dfrac{1}{\sqrt{-g}}\dfrac{\delta S_{EH}}{\delta \Gamma ^\alpha_{\ \mu\nu}}=\dfrac{\partial \mathcal{N}}{\partial C^\alpha_{\ \mu\nu}},
\end{equation}
with $\mathcal{N}=g^{\mu\nu}\mathcal{N}_{\mu\nu}$. If the connection enters the action only algebraically, its equation  may be written as 
\begin{equation}\label{FEPalatini}
P_\alpha^{\ \mu\nu}(C)=\mathcal{B}_\alpha^{\ \mu\nu}(g,\partial\phi,\ldots, (\overset{\scriptscriptstyle g}\nabla\nabla \phi)^{n-1},C,\ldots,C^n).
\end{equation}
From the projective invariance of $\overset{\scriptscriptstyle \Gamma}R$ it follows that\cite{Hehl1978} $P_\alpha^{\ \nu\alpha}=0$ and therefore 
the above  equations are  inconsistent, unless $\mathcal{B}_\alpha^{\ \nu\alpha}=0$.
Assuming the latter condition amounts  to require the projective invariance  of the action $S_C$.
Besides, one can show that the Palatini tensor is 
\begin{equation}\label{Palatini_Def0}
P_{\alpha\mu\nu}=Q_{\alpha \mu\nu}-\dfrac{1}{2}g_{\mu\nu}Q_{\alpha}+g_{\mu\alpha}\Big(\dfrac{1}{2}Q_\nu-\tilde{Q}_\nu\Big)+T_{\mu\alpha\nu}-g_{\mu\nu}T_\alpha+g_{\mu\alpha}T_{\nu},
\end{equation}
where the Weyl vectors are  $Q^\alpha\equiv Q^{\alpha\ \mu}_{\ \mu}$ and $\tilde{Q}_\alpha\equiv Q^\mu _{\ \mu\alpha}$ and the torsion vector is $T_\alpha\equiv T^\beta_{\ \alpha\beta}$. Before going further we note  that  \eqref{Palatini_Def0} can be inverted to express the distortion tensor $C^\alpha_{\ \mu\nu}$ in term of the Palatini tensor:\cite{Iosifidis_2019}
\begin{align*}
C^\alpha_{\ \mu\nu}=\dfrac{1}{2}\left( P^\alpha_{\ \mu\nu}-P_{\mu\nu}^{\ \ \alpha}-P_{\nu\ \mu} ^{\ \alpha}\right)&+\dfrac{1}{4}g_{\mu\nu}\left(P_\lambda^{\ \lambda\alpha}-P^{\alpha  \ \lambda}_{\ \ \lambda}\right) -\dfrac{1}{4}\delta^\alpha_\mu\left(P_{\lambda \  \nu}^{\ \lambda}-P_{\nu\lambda}^{\ \ \lambda}\right)\\
&+\dfrac{1}{4}\delta^{\alpha}_\nu\left(\dfrac{1}{3}P_{\lambda \  \mu}^{\ \lambda}+\dfrac{1}{4}P_{\mu\lambda}^{\ \ \lambda}\right)+\dfrac{1}{3}\delta^{\alpha}_\nu T_\mu.\numberthis\label{C_Palat}
\end{align*}
Therefore  \eqref{FEPalatini} can also be seen as equations for $P_\alpha^{\ \mu\nu}$, while \eqref{C_Palat} expresses  the distortion tensor in term of $P^\alpha_{\ \mu\nu}$ or equivalently in terms of $\mathcal{B}^\alpha_{\ \mu\nu}$, after  using \eqref{FEPalatini}.
The projective symmetry can be used to impose the gauge condition $T_{\mu}=0$.
\section{Construction of the Theory}
In the parametrization of the Generalized Galileon\cite{Kobayashi:2011nu} the action of the metric Horndeski theory contains four arbitrary  functions $G_i(\phi,X)$ ($i=2,3,4,5$)
depending on $\phi$ and on  $X=g^{\alpha\beta}\partial_\beta\phi\partial_\alpha\phi$.
The  action also  contains  the second covariant derivatives of the scalar field 
\begin{equation}\label{SderMetric}
\overset{\scriptscriptstyle(0)}{\Phi}{}^\alpha_{\ \beta}=\overset{\scriptscriptstyle g}\nabla{}^\alpha\overset{\scriptscriptstyle g}\nabla_\beta\phi,
\end{equation}
\noindent and also their powers up to the cubic order $(n=3)$. The theory is determined by the  
Lagrangian  $L_{\rm H}=\sqrt{-g}\left({\cal L}_2+{\cal L}_3+{\cal L}_4+{\cal L}_5\right)$ with
\begin{align*}                  \label{horn}
&{\cal L}_2=G_2,\hspace{0.5cm} {\cal L}_3=G_3\overset{\scriptscriptstyle(0)}{\mathbf{\Phi}}, \hspace{0.5cm}{\cal L}_4=G_4\,\mathcal{R}-2G_{4X}\Big (\overset{\scriptscriptstyle(0)}{\mathbf{\Phi}}{}^2-\big(\overset{\scriptscriptstyle(0)}{\Phi}{}^\alpha_{\ \beta}\big)^2\Big),\numberthis   \\
&{\cal L}_5=G_5 G_{\alpha\beta}\overset{\scriptscriptstyle(0)}{\Phi}{}^{\alpha\beta}
+\dfrac{1}{3} G_{5X}\Big(\overset{\scriptscriptstyle(0)}{\mathbf{\Phi}}{}^3- 3 \overset{\scriptscriptstyle(0)}{\mathbf{\Phi}} \big(\overset{\scriptscriptstyle(0)}{\Phi}{}^\alpha_{\ \beta}\big)^2 + 2\big(\overset{\scriptscriptstyle(0)}{\Phi}{}^\alpha_{\ \beta}\big)^3\Big).          
\end{align*}
Here $\overset{\scriptscriptstyle(0)}{\mathbf{\Phi}}\equiv\overset{\scriptscriptstyle(0)}{\Phi}{}^\alpha_{\ \alpha}=\overset{\scriptscriptstyle g}{\Box}\phi$, $\big(\overset{\scriptscriptstyle(0)}{\Phi}{}^\alpha_{\ \beta}\big)^2=\overset{\scriptscriptstyle(0)}{\Phi}{}^\alpha_{\ \beta}\overset{\scriptscriptstyle(0)}{\Phi}{}_\alpha^{\ \beta}$, similarly for the third power, 
while  $\mathcal{R}=g^{\mu\nu}\mathcal{R}_{\mu\nu}(g)$. Let us stress that 
a metric theory may have various formulations which are equivalent up to a total derivative, but due to the non-metricity and torsion this equivalence may not hold
 in its  MAG versions.  For example the terms $\nabla_\mu\left( G_3\nabla^\mu\phi\right)$ and $\mathcal{R}_{\mu\nu}\nabla^\mu\phi\nabla^\nu\phi-\overset{\scriptscriptstyle(0)}{\mathbf{\Phi}}{}^2 + \big(\overset{\scriptscriptstyle(0)}{\Phi}{}^\alpha_{\ \beta}\big)^2$ are total derivatives in a metric theory but not  in MAG theories. Hence 
{ a} metric-affine formulation of the Horndeski theory { based on a chosen metric parametrization}  will  be just one of many other possible MAG 
 extensions of the theory. 

In the metric-affine context the definition of the second derivatives of the scalar field is not unique because of the non-metricity. 
The possible independent second order covariant derivative operators which reduce to \eqref{SderMetric} in the Levi-Civita limit  are
\begin{align}
&\hspace{5cm}\overset{\scriptscriptstyle(1)}{\Phi}{}^\alpha_{\ \beta}=\overset{\scriptscriptstyle \Gamma}\nabla{}^\alpha\overset{\scriptscriptstyle \Gamma}\nabla_\beta\phi,\label{Operator1}\\
&\overset{\scriptscriptstyle(2)}{\Phi}{}^\alpha_{\ \beta}=g_{\rho\beta}\overset{\scriptscriptstyle \Gamma}\nabla{}^\alpha\overset{\scriptscriptstyle \Gamma}\nabla{}^\rho\phi,\hspace{0.3cm}{\black \overset{\scriptscriptstyle(3)}{\Phi}{}^\alpha_{\ \beta}=\Scale[1.15]{\frac{1}{4}}g_{\mu\nu}\overset{\scriptscriptstyle \Gamma}\nabla{}^\alpha\left(g^{\mu\nu}\overset{\scriptscriptstyle \Gamma}\nabla_\beta\phi\right),\hspace{0.3cm}\overset{\scriptscriptstyle(4)}{\Phi}{}^\alpha_{\ \beta}=\overset{\scriptscriptstyle \Gamma}\nabla_\rho\left(g^{\rho\alpha}\overset{\scriptscriptstyle \Gamma}\nabla_\beta\phi\right),}\label{Operators}
\end{align}
\noindent with the convention $\overset{\scriptscriptstyle \Gamma}\nabla{}^\mu\phi\equiv g^{\mu\nu}\overset{\scriptscriptstyle \Gamma}{\nabla}_\nu\phi$. 
These definitions can be recast in terms of the  ``minimal" operator $\overset{\scriptscriptstyle(1)}{\Phi}{}^\alpha_{\ \beta}=\overset{\scriptscriptstyle(0)}{\Phi}{}^\alpha_{\ \beta}+C^{\mu\alpha}_{\ \ \ \beta}\partial_\mu\phi$ plus  terms  proportional to the non-metricity:
\begin{equation}
\overset{\scriptscriptstyle(2)}{\Phi}{}^\alpha_{\ \beta}=\overset{\scriptscriptstyle(1)}{\Phi}{}^\alpha_{\ \beta}-Q^{\alpha\ \mu}_{\ \beta}\partial_\mu\phi, \qquad
\overset{\scriptscriptstyle(3)}{\Phi}{}^\alpha_{\ \beta}=\overset{\scriptscriptstyle(1)}{\Phi}{}^\alpha_{\ \beta}-\tilde{Q}^{\alpha}\partial_\beta\phi, \qquad \overset{\scriptscriptstyle(4)}{\Phi}{}^\alpha_{\ \beta}=\overset{\scriptscriptstyle(1)}{\Phi}{}^\alpha_{\ \beta}-\frac{1}{4} Q^{\alpha}\partial_\beta\phi.
\end{equation}
Note that the $\overset{\scriptscriptstyle(i)}{\Phi}_{\alpha\beta}=g_{\alpha\sigma}\overset{\scriptscriptstyle(i)}{\Phi}{}^\sigma_{\ \beta}$ tensors are not symmetric, 
hence  one should be careful with the order of indices. 

It seems natural that the MAG versions of the Horndeski theory should respect  the following conditions: 
\begin{romanlist}[(ii)]
	\item $\mathcal{L}_g(g,\partial g,\partial\partial g,\phi,\nabla\phi,\ldots,(\overset{\scriptscriptstyle g}\nabla\nabla \phi)^n)$ must have the structure of the original metric Horndeski theory.\label{itm:1}\\
	\item $\mathcal{L}_{C}(g,\phi,\nabla\phi,\ldots,(\overset{\scriptscriptstyle g}\nabla\nabla \phi)^{n-1},C,\ldots,C^n,\overset{\scriptscriptstyle g}\nabla C)$ must originate from generalized curvature tensors related to the independent connection and from a consistent replacement $\lbrace\overset{\scriptscriptstyle(0)}{\Phi}{}^\alpha_{\ \beta}\rbrace\to\lbrace\overset{\scriptscriptstyle(i)}{\Phi}{}^\alpha_{\ \beta}\rbrace$ in the original metric action. The terms  $\overset{\scriptscriptstyle(i)}{\mathbf{Y}}=\partial_\alpha\phi\partial^\beta\phi \overset{\scriptscriptstyle(i)}\Phi{}^\alpha_{\ \beta}$ may also enter the action. \label{itm:2}
\end{romanlist}
\noindent The most general MAG Lagrangian ${\cal L}_{3}^{\cal MA}$ resulting from the metric-affine extension (\ref{itm:1})-(\ref{itm:2}) of ${\cal L}_3$ is constructed from  
$\Bigl\lbrace\hphantom{i}\overset{\scriptscriptstyle(1)}{\mathbf{\Phi}}$, $\overset{\scriptscriptstyle(2)}{\mathbf{\Phi}}$, $\overset{\scriptscriptstyle(3)}{\mathbf{\Phi}}$, $\overset{\scriptscriptstyle(1)}{\mathbf{Y}}$, $\overset{\scriptscriptstyle(2)}{\mathbf{Y}}\Big\rbrace$ \footnote{{\black The traces and contractions with the derivatives of the scalar field of the operators defined in \eqref{Operator1}-\eqref{Operators} are not all independent. One has for example $\hphantom{i}\overset{\scriptscriptstyle(2)}{\mathbf{\Phi}}=\hphantom{i}\overset{\scriptscriptstyle(4)}{\mathbf{\Phi}}$.} } and can be written after integration by part as 
\begin{equation}\label{Rep3b}
\mathcal{L}_{3}^{\cal MA}=G_3(\phi,X)\overset{\scriptscriptstyle(0)}{\mathbf{\Phi}}+C_1 Q^\alpha\partial_\alpha\phi+C_2\tilde{Q}^\alpha\partial_\alpha\phi
+C_3 Q^\alpha_{\ \mu\nu}\partial_\alpha\phi\,\partial^\mu\phi\,\partial^\nu\phi+C_4\partial_\alpha\phi\, T^\alpha,
\end{equation} 

\noindent where the $C_i$ are functions of $\phi$ and $X$. Within the $n=1$ Horndeski class, the generalized $\mathcal{L}_4$ part compatible with our requirements (\ref{itm:1})-(\ref{itm:2}) is obtained by setting in \eqref{horn} $G_{4}=G_4(\phi)$, hence 
\begin{equation}\label{Rep4a}
\mathcal{L}_{4}^{\cal MA}=G_4(\phi)\,\overset{\scriptscriptstyle \Gamma}R
\end{equation}
with $\overset{\scriptscriptstyle \Gamma}R=g^{\mu\nu}\overset{\scriptscriptstyle \Gamma}R_{\mu\nu}$. Adding
the $G_5$ terms would be incompatible with our assumptions. 
As a result, 
including also the ${\cal L}_2$ term, yields  the MAG  action 
\begin{align}\label{actionST}
S^{\cal MA}=\int d^4x\sqrt{-g}\Big(&\mathcal{L}_{4}^{\cal MA}+\mathcal{L}_{3}^{\cal MA}+{\cal L}_2\Bigr).
\end{align} 
{\black A subset of this theory build solely upon the scalars $\overset{\scriptscriptstyle(1)}{\mathbf{\Phi}}$ and $\overset{\scriptscriptstyle(2)}{\mathbf{\Phi}}$ was studied in Ref. \citen{Shimada:2018lnm} for particular models in the context of inflation.} 
\section{Solution for the Connection}
Varying \eqref{actionST} with respect to $C^\alpha_{\ \mu\nu}$ yields 
\begin{equation}\label{EqPalatini}
P_{\alpha\mu\nu}=\mathcal{B}_{\alpha\mu\nu}
\end{equation}
where 
 \begin{equation}
 \mathcal{B}_{\alpha\mu\nu}=\dfrac{1}{G_4}\left(\Delta_{\alpha\mu\nu}+ G_{4\phi}( g_{\mu\nu}  \nabla_\alpha\phi- g_{\alpha\mu}\nabla_\nu\phi)\right)
\end{equation}
with 
\begin{eqnarray}\label{HM0}
\Delta_{\alpha\mu\nu}&=&C_2\, g_{\mu\nu}\nabla_\alpha\phi+\left(2C_1-C_4\right) g_{\alpha\nu}\nabla_\mu\phi \nonumber \\
&&+\left(C_2+C_4\right)g_{\alpha\mu}\nabla_\nu\phi+2C_3\nabla_\alpha\phi\nabla_\mu\phi\nabla_\nu\phi.
\end{eqnarray}
The projective invariance condition $\mathcal{B}_\alpha^{\ \nu\alpha}=0$ imposes $C_1=-\frac{1}{4}\big(C_2+C_3 {\black X}-\frac{3}{2}C_4\big)$ and
the solution for the distortion tensor is then obtained by injecting \eqref{EqPalatini} in \eqref{C_Palat}:
\begin{align*}
C^\alpha_{\ \mu\nu}=&-\dfrac{1}{4G_4}\Big(4C_3\nabla^{\alpha}\phi\nabla_{\mu}\phi\nabla_{\nu}\phi+g_{\mu\nu}\nabla^\alpha\phi\big(2G_{4\phi}-C_2-C_3 X-\dfrac{3}{2}C_4\big)\\
&-2\delta^{\alpha}_{(\mu}\nabla_{\nu)}\phi\big(2G_{4\phi}+C_2+C_3 X-\dfrac{1}{2}C_4\big)\Big)+\dfrac{1}{3}\delta^{\alpha}_{\nu} T_\mu.\numberthis\label{DistortionSolution}
\end{align*}
A particularly interesting subset of solutions corresponds to the choice $C_3=0$ and $C_4=-2C_2$. In this class of theories,
setting\footnote{There exist other gauge choices such that the connection takes the generalized Weyl form.\cite{Jimenez:2015fva}} $T_\mu=0$, the solution for the distortion tensor reduces to 
\begin{equation}\label{WeylConnection}
C^\alpha_{\ \mu\nu}=-g_{\mu\nu}A^\alpha+2\delta^{\alpha}_{(\mu}A_{\nu)} \quad \text{with} \quad A_{\mu}=\dfrac{1}{2G_4}\left(G_{4\phi}+C_2\right)\partial_\mu\phi,
\end{equation}
and the non-metricity  is 
\begin{equation}\label{WeylComp}
Q_{\alpha\mu\nu}\equiv \overset{\scriptscriptstyle \Gamma}\nabla_\alpha g_{\mu\nu}=-2A_\alpha g_{\mu\nu}.
\end{equation}
A geometry defined by a family of conformally related pseudo-Riemannian metrics with a connection respecting the compability condition \eqref{WeylComp} is called  Weyl geometry \cite{Scholz:2011za,WeylGeoHis}. The connection $\Gamma^\alpha_{~\mu\nu}$  is invariant under local Weyl transformations 
\begin{equation}
g_{\mu\nu}\to \bar{g}_{\mu\nu}=e^{2\Lambda (x)}g_{\mu\nu}, \quad \text{and} \quad A_\alpha\to \bar{A}_\alpha= A_\alpha-\partial_\alpha \Lambda (x).
\end{equation}
The metric-affine  theory respecting these conditions is 
\begin{equation}\label{WeylAction}
S_{\text{Weyl}}=\int d^4 x \sqrt{-g}\Big(\mathcal{L}_H+G_4\, \mathcal{N}+C_2\left(\tilde{Q}^\alpha-Q^\alpha-2T^\alpha\right)\partial_\alpha\phi\Bigr), 
\end{equation}
where $\mathcal{L}_H$ is the Levi-Civita limit of 
$\mathcal{L}_{4}^{\cal MA}+\mathcal{L}_{3}^{\cal MA}+{\cal L}_2$ in \eqref{actionST}.
When $C_2$ vanishes then the 1-form $A_\mu=\dfrac{1}{2}\partial_\mu \ln G_4(\phi)$ is exact. 
A Weyl geometry where $A_\mu$ is  exact is called {\it Weyl integrable}. This subset of Weyl geometries  is intimately related to the existence of a covariantly constant conformally related metric\cite{Romero_2012},
\begin{equation}
\tilde{g}_{\mu\nu}=G_4(\phi)g_{\mu\nu}, \qquad \overset{\scriptscriptstyle \Gamma}\nabla_\alpha \tilde{g}_{\mu\nu}=0.
\end{equation}
\noindent In other words, the connection $\Gamma^\alpha_{~\mu\nu}$ is totally determined by the metric $\tilde{g}$, in which case the latter 
is said to define the  Riemannian frame. On the contrary, if $C_2\neq 0$ then 
$A_\mu$ is not exact and  the connection $\Gamma^\alpha_{~\mu\nu}$ is not Levi-Civita.  Still, 
as we saw previously, the theory admits a dynamically equivalent effective metric description. 
\section{Dynamically Equivalent Metric Theory}
 Integrating the connection out from  \eqref{actionST} via setting it to the value 
\eqref{DistortionSolution}, 
we obtain the Lagrangian of the dynamically equivalent metric theory, 
\begin{equation}\label{Leff}
\tilde{L}_H=\sqrt{-g}\left(\mathcal{L}_H+K_{eff}(\phi,X)\right),
\end{equation}
where  the  K-essence  term is 
\begin{equation}\label{Keff}
K_{eff}=\dfrac{X}{32 G_4}\Big(\big(4XC_3\big)^2-9\big(2C_2+C_4+\dfrac{2}{3}XC_3\big)^2+3\big(2C_4-4G_{4\phi}\big)^2\Big).
\end{equation}
This Lagrangian actually belongs to  the Horndeski family, hence 
the  theory is free from the Ostrogradsky ghost. If the connection assumes the Weyl form \eqref{WeylConnection} then one has 
\begin{equation}
K_{eff}=\dfrac{3X}{2G_4}\left(C_2+G_{4\phi}\right)^2.
\end{equation}
Remarkably, there is a non trivial yet simple limit where $K_{eff}$ goes to zero:
\begin{equation}\label{LimitK}
C_2\to -XC_3-G_{4\phi}, \hspace{0.4cm} C_4\to 2 G_{4\phi},\quad \Rightarrow\quad C^\alpha_{\ \mu\nu}\to-\dfrac{1}{G_4}C_3\nabla^{\alpha}\phi\nabla_{\mu}\phi\nabla_{\nu}\phi.
\end{equation}
The total equivalence between the effective metric theory and the original metric Horndeski theory in this limit (${\tilde{L}}_H={L}_H$) results from the fact that for any distortion tensor of the form $C^\alpha_{\ \mu\nu}=\xi^\alpha\, \xi_\mu\,  \xi_\nu$, where $\xi$ is an arbitrary vector fields, the non-Riemannian part of the Ricci scalar vanishes: $\mathcal{N}=0$.

\section{Conclusion}
We analysed the metric-affine scalar-tensor theory linear in second derivatives of the scalar field obtained by 
 relaxing the minimal derivative coupling prescription. {\black The full theory  has never been studied before, and it may have 
 interesting applications, for example in cosmology \cite{Helpin:2019kcq,Shimada:2018lnm}.}
It could be interesting to see how the introduction of all the $\overset{\scriptscriptstyle(i)}{\Phi}{}^\alpha_{\ \beta}$ would impact the results 
of Refs. \citen{Aoki:2018lwx} and \citen{Shimada:2018lnm}.

\section*{Acknowledgements}
The work of M.S.V. was  partly supported by the  PRC CNRS/RFBR grant and also 
by the Russian Government Program of Competitive Growth 
of the Kazan Federal University.


\begin{thebibliography}{10}
\expandafter\ifx\csname urlstyle\endcsname\relax
  \providecommand{\doi}[1]{doi:\discretionary{}{}{}#1}\else
  \providecommand{\doi}{doi:\discretionary{}{}{}\begingroup
  \urlstyle{rm}\Url}\fi

\bibitem{Horndeski1974}
G.~W. Horndeski, {\it Int. J. Theor. Phys.} {\bf 10},
  363 (1974).

\bibitem{PhysRevD.89.064046}
M.~Zumalac\'arregui and J.~Garc\'{\i}a-Bellido, {\it Phys. Rev. D} {\bf 89},
  064046 (2014). 

\bibitem{Gleyzes:2014dya}
J.~Gleyzes, D.~Langlois, F.~Piazza and F.~Vernizzi, {\it Phys. Rev. Lett.} {\bf
  114},   211101  (2015).

\bibitem{Langlois_2016}
D.~Langlois and K.~Noui, {\it JCAP}
  {\bf 2016}, 034 (2016). 

\bibitem{BenAchour:2016fzp}
J.~Ben~Achour, M.~Crisostomi, K.~Koyama, D.~Langlois, K.~Noui and G.~Tasinato,
  {\it JHEP} {\bf 12},   100  (2016).
 
\bibitem{Woodard:2015zca}
R.~P. Woodard, {\it Scholarpedia} {\bf 10},   32243  (2015).

\bibitem{Nicolis:2008in}
A.~Nicolis, R.~Rattazzi and E.~Trincherini, {\it Phys. Rev. D} {\bf 79},
  064036  (2009).
  
\bibitem{Deffayet:2009wt}
C.~Deffayet, G.~Esposito-Farese and A.~Vikman, {\it Phys. Rev. D} {\bf 79},
  084003  (2009). 

\bibitem{Deffayet:2011gz}
C.~Deffayet, X.~Gao, D.~A. Steer and G.~Zahariade, {\it Phys. Rev. D} {\bf 84},
   064039  (2011). 

\bibitem{Hehl1978}
F.~W. Hehl and G.~D. Kerlick, {\it Gen. Rel. Grav.} {\bf 9},
  691 (1978). 

\bibitem{Li_2012}
M.~Li and X.~Wang, {\it JCAP} {\bf
  2012}, 010 (2012). 

\bibitem{PhysRevD.100.064018}
S.~Bahamonde, K.~F. Dialektopoulos and J.~L. Said, {\it Phys. Rev. D} {\bf
  100},   064018 (2019). 

\bibitem{GALTSOV2019453}
D.~Gal'tsov and S.~Zhidkova, {\it Phys. Lett. B} {\bf 790}, 453   (2019). 
 
\bibitem{Helpin:2019kcq}
T.~Helpin and M.~S. Volkov (2019), {{\ttfamily arXiv:1906.07607
		[hep-th]}}

\bibitem{PhysRevD.96.084023}
J.~Barrientos, F.~Cordonier-Tello, F.~Izaurieta, P.~Medina, D.~Narbona,
  E.~Rodr\'{\i}guez and O.~Valdivia, {\it Phys. Rev. D} {\bf 96},   084023 (2017).

\bibitem{Aoki:2018lwx}
K.~Aoki and K.~Shimada, {\it Phys. Rev. D} {\bf 98},   044038  (2018).
 
\bibitem{PhysRevD.99.104020}
K.~Shimada, K.~Aoki and K.-i. Maeda, {\it Phys. Rev. D} {\bf 99},   104020 (2019). 

\bibitem{PhysRevD.100.044037}
K.~Aoki and K.~Shimada, {\it Phys. Rev. D} {\bf 100},   044037 (2019). 

\bibitem{Shimada:2018lnm}
K.~Shimada, K.~Aoki and K.-i. Maeda, {\em Phys. Rev.} {\bf D99},   104020
(2019).
	
\bibitem{Iosifidis_2019}
D.~Iosifidis, {\em Classical and Quantum Gravity} {\bf 36},   085001 (2019).

\bibitem{Kobayashi:2011nu}
T.~Kobayashi, M.~Yamaguchi and J.~Yokoyama, {\it Prog. Theor. Phys.} {\bf 126},
  511  (2011). 

\bibitem{Jimenez:2015fva}
J.~Beltran~Jimenez and T.~S. Koivisto, {\em Phys. Lett. B} {\bf 756}, 400
  (2016).
  
\bibitem{Scholz:2011za}
E.~Scholz (2011), {{\ttfamily
		arXiv:1111.3220 [math.HO]}}.

\bibitem{WeylGeoHis}
L.~O'Raifeartaigh and N.~Straumann, {\it Rev. Mod. Phys.} {\bf 72}, 1 (2000). 

\bibitem{Romero_2012}
C.~Romero, J.~B. Fonseca-Neto and M.~L. Pucheu, {\it Class. Quant.
  Grav.} {\bf 29},   155015 (2012). 

\end{thebibliography}

\end{document}